\documentclass[a4paper,fleqn,usenatbib,useAMS]{mnras}

\usepackage{tikz,xcolor,hyperref}

\definecolor{lime}{HTML}{A6CE39}
\DeclareRobustCommand{\orcidicon}{
	\begin{tikzpicture}
	\draw[lime, fill=lime] (0,0) 
	circle [radius=0.13] 
	node[white] {{\fontfamily{qag}\selectfont \tiny ID}};
	\draw[white, fill=white] (-0.0625,0.095) 
	circle [radius=0.007];
	\end{tikzpicture}
	\hspace{-2mm}
}

\foreach \x in {A, ..., Z}{\expandafter\xdef\csname orcid\x\endcsname{\noexpand\href{https://orcid.org/\csname orcidauthor\x\endcsname}
			{\noexpand\orcidicon}}
}

\usepackage{graphicx}	
\usepackage{amsmath}	
\usepackage{amssymb}	
\usepackage{multicol}        
\usepackage{bm}		
\usepackage{pdflscape}	
\usepackage[encapsulated]{CJK}
\usepackage{ucs}
\usepackage[utf8x]{inputenc}
\usepackage{multirow}
\usepackage{booktabs,caption}
\usepackage[flushleft]{threeparttable}

\usepackage[T1]{fontenc}
\usepackage{ae,aecompl}

\usepackage{newtxtext,newtxmath}

\title[Spatio-kinematic Models of Nova Remnants]{
Spatio-kinematic models of five nova remnants: correlations between nova shell axial ratio, expansion velocity, and speed class
}

\author[Santamar\'\i a et al.]
{E.\ Santamar\'{i}a$^{1,2\thanks{E-mail: {\bf edgar.santamaria8808@alumnos.udg.mx}}\orcidA{}}$,
M.A.\,Guerrero$^{3\orcidB{}}$,
S.\,Zavala$^{4\orcidC}$,
G.\,Ramos-Larios$^{1,2\orcidD{}}$,
J.A.\,Toal\'{a}$^{5\orcidE{}}$
and L.\,Sabin$^{6\orcidF{}}$
\\
$^1$Universidad de Guadalajara, CUCEI, Blvd. Marcelino Garc\'\i a Barrag\'an 1421, 
44430, Guadalajara, Jalisco, Mexico \\
$^2$Instituto de Astronom\'\i a y Meteorolog\'\i a, Dpto.\ de F\'\i sica,
CUCEI, Av.\ Vallarta 2602, 44130, Guadalajara, Jalisco, Mexico\\
$^3$Instituto de Astrof\'\i sica de Andaluc\'\i a, IAA-CSIC, Glorieta de la
Astronom\'\i a s/n, 18008, Granada, Spain \\
$^4$Tecnol\'ogico Nacional de M\'exico / I. T. Ensenada, Depto. de Ingenier\'ia El\'ectrica y Electr\'onica, Ensenada, B.C., Mexico \\
$^5$Instituto de Radioastronom\'{i}a y Astrof\'{i}sica (IRyA), UNAM Campus Morelia,
Apartado postal 3-72, 58090 Morelia, Michoac\'{a}n, Mexico\\
$^6$Instituto de Astronom\'\i a, Universidad Nacional Aut\'onoma de M\'exico,
Apdo.\ Postal 877, C.P. 22860, Ensenada, B.C., Mexico \\
}

\pubyear{2022}

\begin{document}
\label{firstpage}
\pagerange{\pageref{firstpage}--\pageref{lastpage}}
\maketitle


\begin{abstract}
We present long-slit intermediate-dispersion spectroscopic observations and narrow-band direct imaging of four classical nova shells, namely T\,Aur, HR\,Del, DQ\,Her and QU\,Vul, and the nova-like source CK\,Vul. 
These are used to construct models of their nebular remnants using the morpho-kinematic modelling tool {\sc shape} to reveal their 3D shape.
All these nova remnants but CK\,Vul can be described by prolate ellipsoidal shells with different eccentricity degree, from the spherical QU\,Vul to the highly elongated shell with an equatorial component HR\,Del.  
On the other hand, CK\,Vul shows a more complex structure, with two pairs of nested bipolar lobes. 
The spatio-kinematic properties of the ellipsoidal nova shells derived from our models include their true axial ratios. This parameter is expected to correlate with the expansion velocity and decline time $t_{3}$ (i.e., their speed class) of a nova as the result the interaction of the ejecta with the circumstellar material and rotation speed and magnetic field of the white dwarf. We have compared these three parameters including data available in the literature for another two nova shells, V533\,Her and FH\,Ser.  
There is an anti-correlation between the expansion velocity and the axial ratio and decline time $t_{3}$ for nova remnants with ellipsoidal morphology, and a correlation between their axial ratios and decline times $t_{3}$, confirming theoretical expectations that the fastest expanding novae have the smallest axial ratios. We note that the high expansion velocity of the nova shell HR\,Del of 615 km~s$^{-1}$ is inconsistent with its long decline time $t_3$ of 250 days. 
\end{abstract}




\section{Introduction}

Classical novae (CNe) are sudden explosive mass ejection, in which a red dwarf or red giant transfers hydrogen-rich material to a carbon-oxygen (CO) or oxygen-neon (ONe) white dwarf (WD) via an accretion disk with average accretion rate $\sim$1.3 $\times 10^{-10}$ M$_{\odot}$ yr$^{-1}$ \citep{2018ApJ...860..110S}. 
The process continues until the material accreted onto the WD reaches a critical mass limit, when a thermonuclear runaway occurs, achieving surface temperatures $\sim$10$^{8}$ K in a few hundred seconds \citep{2016PASP..128e1001S}. 
During this process, a large amount of energy ($\approx$10$^{44}$ erg) is released and distributed in the accretion disk, causing its expansion and violent ejection from the system. The ejected mass amounts to 10$^{-5}$--10$^{-4}$ M$_{\odot}$ \citep{1998PASP..110....3G} with velocities ranging from 100 to 1000 km~s$^{-1}$ \citep{2010AN....331..160B}. After this event, the ejected mass will expand forming a nova shell or remnant that will eventually mix with the interstellar medium (ISM).

The morphology of a nova remnant can be attributed to different mechanisms acting during this eruptive event. The common envelope phase in the early stage during the outburst is accepted as one of the most important shaping agent. The stellar companion can be expected to contribute significantly to the angular momentum of the system during this phase, which may bring important consequences for the nova shell shaping \citep{LIVIO1990}. The mass loss from the secondary star can be highly concentrated on the orbital plane to produce a bipolar structure during the nova event \citep{2013ASPC..469..323M}. Other physical processes can also affect the nova shaping, including the rotation speed of the WD \citep{PORTER1998}, its magnetic field, and interactions of the ejected material with the stellar companion and accretion disk. Jets might also play a crucial role \citep[see, e.g.,][]{Montez2021}. Since the effects of all these mechanisms in the nova shaping are expected to vary for novae of different velocity classes \citep{LLOYD1997}, the larger asphericities of the remnants of slow novae relative to those of fast novae \citep{1995MNRAS.276..353S} might probe their relative importance.

\begin{table*}
\centering
\setlength{\columnwidth}{0.1\columnwidth}
\setlength{\tabcolsep}{2.58\tabcolsep}
\caption{Properties of the objects in this work.}
\label{tab:nov}
\begin{tabular}{lclclcrc}
\hline
\hline

\multicolumn{1}{c}{Object} &
\multicolumn{1}{c}{$l,b$} & 
\multicolumn{1}{c}{Nova Type} &
\multicolumn{1}{c}{$t_{3}$} &
\multicolumn{1}{c}{Outburst Date} & 
\multicolumn{1}{c}{Distance} & 
\multicolumn{1}{c}{$z$} &
\multicolumn{1}{c}{References} \\ 

\multicolumn{1}{c}{} &
\multicolumn{1}{c}{($^\circ$)} & 
\multicolumn{1}{c}{} & 
\multicolumn{1}{c}{(days)} & 
\multicolumn{1}{c}{} & 
\multicolumn{1}{c}{(pc)} & 
\multicolumn{1}{c}{(pc)} & 
\multicolumn{1}{c}{} \\ 
\hline

T\,Aur  & 177.14$-$1.70  & moderately fast    & 100 & ~~1891 Dec & ~~$840^{+29}_{-27}$   & 25 & 1 \\
HR\,Del    &  63.43$-$13.97 & slow    & 250 & ~~1967 Jul &  893$^{+16}_{-16}$ & $231$ & 1 \\
DQ\,Her   &  73.15$+$26.44 & moderately fast   &  94 & ~~1934 Dec & ~~495$^{+5}_{-5}$    & 220 & 1 \\
CK\,Vul    & 63.38$+$0.18  & moderately fast  & 100 & ~~1670 Jun & ~~3200$^{+900}_{-600}$   & 77 & 2 \\
QU\,Vul   &  68.51$-$6.02  & fast   &  49 & ~~1984 Dec &  900$^{+350}_{-195}$ & $187$ & 1 \\

\hline
\end{tabular}
\vspace{0.15cm}
{\parbox{6.44in}{
\footnotesize \textbf{Note}: References to distances were obtained from (1) \textit{Gaia} Early Data Release 3 (EDR3) \citep{2021AJ....161..147B} and (2) \cite{2020ApJ...904L..23B}.}
}
\end{table*}

\begin{table*}
\centering
\setlength{\columnwidth}{0.1\columnwidth}
\setlength{\tabcolsep}{3.73\tabcolsep}
\caption{Details of the imaging observations. \label{tab:obs}}
\begin{tabular}{llllrc}
\hline
\hline

\multicolumn{1}{c}{Object} &
\multicolumn{1}{c}{Date} &
\multicolumn{1}{c}{Telescope and} & 
\multicolumn{1}{c}{Main Filter} & 
\multicolumn{1}{c}{Exposure} &
\multicolumn{1}{c}{Spatial} \\ 

\multicolumn{1}{c}{} &
\multicolumn{1}{c}{} & 
\multicolumn{1}{c}{Instrument} & 
\multicolumn{1}{c}{} & 
\multicolumn{1}{c}{Time} &
\multicolumn{1}{c}{Resolution} \\ 

\multicolumn{1}{c}{} &
\multicolumn{1}{c}{} & 
\multicolumn{1}{c}{} & 
\multicolumn{1}{c}{} &
\multicolumn{1}{c}{(s)} & 
\multicolumn{1}{c}{($^{\prime\prime}$)} \\ 
\hline

T\,Aur  & 2016 Nov 27 & NOT ALFOSC    & NOT \#21 H$\alpha$ & 1800~~~ & 0.9 \\
HR\,Del & 2020 Jul 27 & NOT ALFOSC    & NOT \#64 H$\alpha$ &  900~~~ & 0.8 \\
DQ\,Her & 2017 May 27 & NOT ALFOSC    & OSN H01 H$\alpha$  & 2700~~~ & 0.8 \\ 
CK\,Vul & 2018 Jun 7  & NOT ALFOSC    & OSN H01 H$\alpha$  & 2700~~~ & 1.5 \\
        & 2010 Jun 22 & GEMINI-N GMOS & H$\alpha$\_G0310   &  300~~~ & 1.0 \\
QU\,Vul	& 2020 Jul 27 & NOT ALFOSC    & NOT \#64 H$\alpha$ & 4800~~~ & 0.7  \\
\hline
\end{tabular}
\vspace{0.15cm}
\end{table*}

Detailed 3D models of the ejected shells can be achieved combining high-resolution spectroscopic observations and optical images to determine the spatio-kinematical properties of the structural components. These provide a full description of a nova remnant, including its expansion parallax and expansion velocity for the determination of its distance, 3D shape, inclination angle, ejected mass, and kinetic energy. This information can be used to assess the effects of the different processes involved in nova shaping, which are also present in the formation of planetary nebulae (PNe) or the evolution of supernova remnants (SNRs). Processes of shocks interactions and mix into the ISM can also be investigated. Understanding the 3D shape of a nova remnant thus provides important clues about the nature of the nova outburst. 

\begin{table*}
\centering
\setlength{\columnwidth}{0.1\columnwidth}
\setlength{\tabcolsep}{2.34\tabcolsep}
\caption{Spectroscopic observing log. \label{tab:slits}}
\begin{tabular}{llcccrlc}
\hline
\hline

\multicolumn{1}{c}{Object} &
\multicolumn{1}{c}{Telescope and Instrument} &
\multicolumn{1}{c}{Spectral Range} &
\multicolumn{1}{c}{Slit Width} &
\multicolumn{1}{c}{Slits} & 
\multicolumn{1}{c}{PA} &
\multicolumn{1}{c}{Date} &
\multicolumn{1}{c}{Exposure} \\ 

\multicolumn{1}{c}{} &
\multicolumn{1}{c}{} & 
\multicolumn{1}{c}{(\AA)} &
\multicolumn{1}{c}{($^{\prime\prime}$)} &
\multicolumn{1}{c}{} & 
\multicolumn{1}{c}{($^{\circ}$)} &
\multicolumn{1}{c}{} &
\multicolumn{1}{c}{Time (s)} \\ 

\hline
T\,Aur  & ~~~~NOT ALFOSC  & 6330--6870 & 0.9 & S1  & 65 & 2020 Jan 24 & 5400 \\
        & ~~~~NOT ALFOSC  & 6330--6870 & 0.9 & S2  & 150 & 2020 Jan 25 & 5400 \\
HR\,Del & ~~~~NOT ALFOSC  & 6330--6870 & 0.75 & S1  & 45 & 2018 Jun 8 & 1350 \\
        & ~~~~NOT ALFOSC  & 6330--6870 & 0.5 & S2  & 135 & 2020 Jul 28 & 1800 \\
DQ\,Her & ~~~~OAN-SPM 2.1m MES & 6545--6590 & 1.9 & S1  & 0 & 2018 Jul 23  & 1200 \\
        & ~~~~OAN-SPM 2.1m MES & 6545--6590 & 1.9 & S2  & 315 & 2018 May 6 & 1800 \\
CK\,Vul & ~~~~NOT ALFOSC  & 6330--6870 & 0.9 & S1  & 174 & 2020 Jul 26  & 3600 \\
        & ~~~~NOT ALFOSC  & 3650--7110 & 0.5 & S2  & 88 & 2021 Jun 13  & 2700 \\
QU\,Vul	& ~~~~NOT ALFOSC  & 6330--6870 & 0.5 & S1    & 123 & 2020 Jul 29 & 3000 \\
\hline
\end{tabular}
\vspace{0.15cm}
\end{table*}

Despite the importance of the determination of the nova ejecta spatio-kinematic properties, only a handful of such studies are available. 
Most of them report direct images and spectroscopic observations through a single long-slit to derive expansion rates and distances \citep[e.g., in FH\,Ser and QU\,Vul,][]{VAL1997}. 
\cite{2003MNRAS.344.1219H} show a combination of long-slit spectra and optical images of HR\,Del in order to produce a kinematic model, concluding that the remnant can be represented by an ellipsoidal structure. \cite{VOR2007} suggested a collimated stellar wind along the polar direction of DQ\,Her that accelerates fragments of material from clumps to velocities up to $\sim$800-900 km s$^{-1}$. 
This polar wind could be the result of the interaction of the present stellar wind with the circumstellar material, but also collimated by magnetic fields or by the rotation of the WD. Adding complexity to the shaping mechanism of DQ\,Her, \citet{2020MNRAS.495.4372T} discovered recently the presence of a magnetised jet, but projected on the plane of the sky perpendicularly to the major nebular axis. 
In the work of \citet{2013MNRAS.432..167H} on CK\,Vul, a range of ejection velocities was found. The nature of these ejecta and the decrease of the nebular flux in the last twenty years was discussed by the authors. As the sophistication of the observations increases, more and more interesting features are revealed. \cite{2013MNRAS.433.1991R} show a morpho-kinematic model of KT\,Eri and determined its 3D shape to be similar to that of a dumbbell structure with expansion velocities up to $\sim$2800 km s$^{-1}$. \cite{2017MNRAS.465..739S} built a 3D model of the distribution of clumps in AT\,Cnc and concluded that the ejection occurred $\simeq$330 yr ago, which is consistent with the cooling time of a WD after a CNe eruption and the predicted scale for the hibernation time.

In this work we present a kinematic analysis using intermediate- and high-dispersion long-slit spectra and narrow-band images to produce morpho-kinematic models of 4 nova shells (namely T\,Aur, HR\,Del, DQ\,Her and QU\,Vul) and the nova-like object CK\,Vul (see Tab.~\ref{tab:nov} for basic information on these sources).
To reproduce simultaneously the morphology and kinematics, we used the interactive 3D {\sc{shape}} software \citep{2011ITVCG..17..454S,Steffen2012}. 
The observations are presented in Section \ref{2}, 
the kinematics and the spatio-kinematic models are described in Section \ref{3}, and 
a discussion is presented in Section \ref{4}. 
A final summary is presented in Section \ref{5}.

\begin{figure*}
\begin{center}
\includegraphics[width=0.95\linewidth]{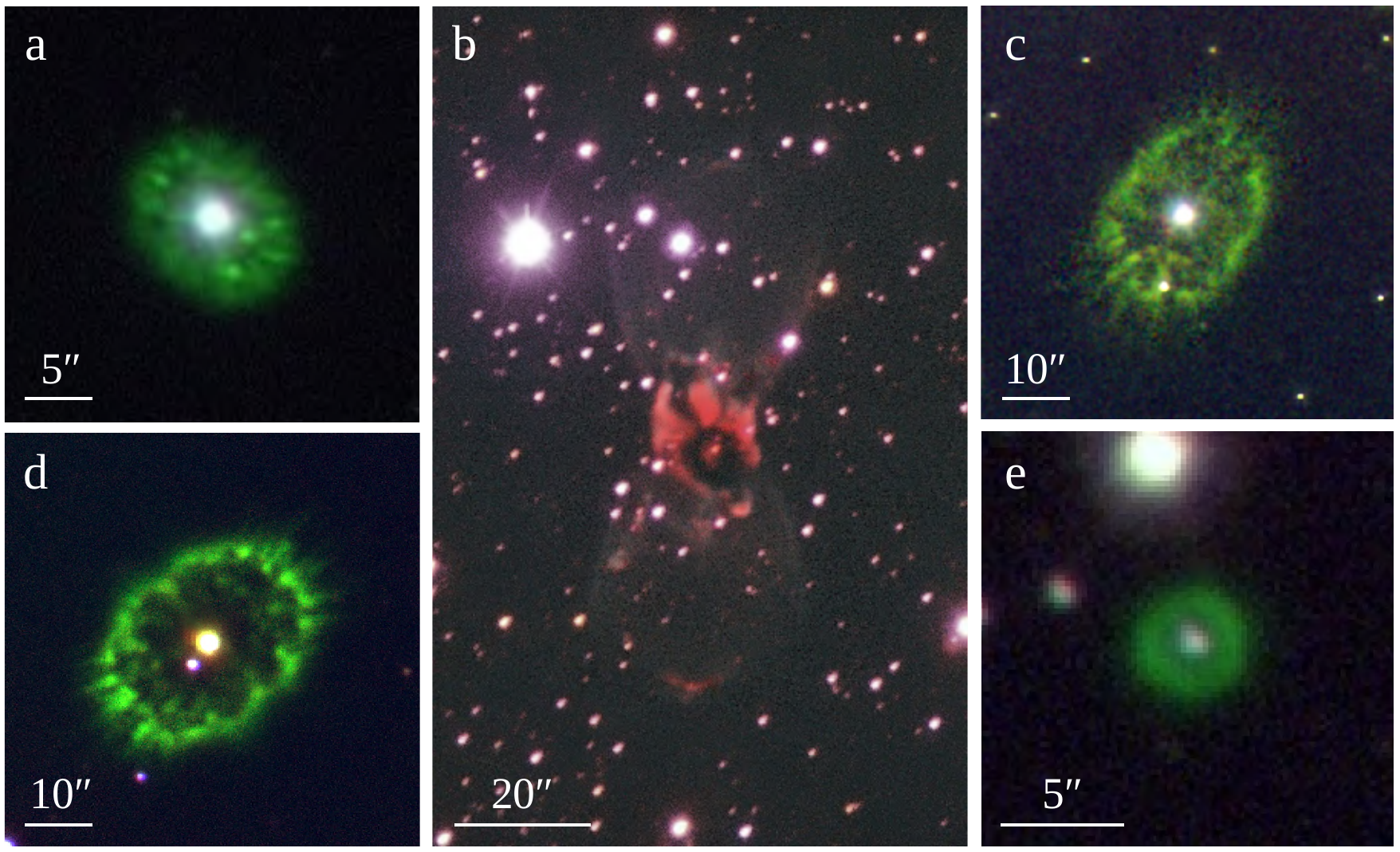}
\caption{Colour-composite images of the nova shells sources. a) HR\,Del, c) T\,Aur, d) DQ\,Her, e) QU\,Vul and the nova-like b) CK\,Vul respectively. The \textit{a} and \textit{e} panels are composites of 
broad-band $g'$ SDSS $\lambda$4800 (blue), narrow-band H$\alpha$ 
$\lambda$6563 (green) and $r'$ SDSS $\lambda$6180 filter for the red color. For the \textit{c} and \textit{d} panels, were used the $g'$ SDSS $\lambda$4800 (blue), H$\alpha$ 
$\lambda$6563 (green) and [N~{\sc ii}] $\lambda$6583 (red) filters, whilst the panel \textit{b}, is the combination of $g'$ SDSS $\lambda$4800 (blue), $r'$ SDSS $\lambda$6180 (green) and H$\alpha$ 
$\lambda$6563 (red) filters plus, the use of the H$\alpha$ filter from the GEMINI-GMOS camera. These particular combination of filters have made it possible to highlight the weak nebular bipolar emission. For all the images, North is up and East to the left.}
\label{fig:nov}
\end{center}
\end{figure*}

\section{Observations}\label{2}

\subsection{Optical imaging}

The optical images of the sample of novae presented in this work are described in Table~\ref{tab:obs}. 
The images were acquired using the Alhambra Faint Object Spectrograph and Camera (ALFOSC)\footnote{\url{http://www.not.iac.es/instruments/alfosc/}} installed at the Cassegrain focus of the 2.5m Nordic Optical Telescope (NOT) at the Roque de los Muchachos Observatory (ORM, La Palma, Spain). 
The detector was an E2V 231-42 2k$\times$2k CCD with a pixel size of 15.0 $\mu$m, resulting in a plate scale of 0.214 arcsec pix$^{-1}$ and a field of view (FoV) of 6.4$\times$6.4 arcmin. 
The images were obtained through the OSN (Sierra Nevada Observatory) H$\alpha$ filter H01 ($\lambda_{\rm c}=6563$\,{\AA}, FWHM $=$ 13\,{\AA}) and the NOT filters H$\alpha$ \#64 ($\lambda_{\rm c}=6562$\,{\AA}, FWHM $=$ 46\,{\AA}) and H$\alpha$ \#21 ($\lambda_{\rm c}=6564$\,{\AA}, FWHM $=$ 33\,{\AA}). 

Total exposure times and spatial resolutions, as determined from the FWHM of field stars, are listed in Table~\ref{tab:obs}. 
Images were also acquired in other filters: OSN E16 [N~{\sc ii}] ($\lambda_{\rm c}=6584$\,{\AA}, FWHM $=$ 10\,{\AA}) and NOT $g'$ SDSS \#83 ($\lambda_{\rm c}=4800$\,{\AA}, FWHM $=$ 1450\,{\AA}) and $r'$ SDSS \#110 ($\lambda_{\rm c}=6180$\,{\AA}, FWHM $=$ 1480\,{\AA}) filters, as described briefly in the caption of Figure \ref{fig:nov}, to obtain the color-composite pictures of our sample presented in that figure.

The data were processed following standard {\sc iraf}\footnote{{\sc iraf} is distributed by the National Optical Astronomy Observatory, which is operated by the Association of Universities for Research in Astronomy (AURA) under a cooperative agreement with the National Science Foundation.} \citep{1993ASPC...52..173T} routines, including bias subtraction and flat field correction using appropriate sky flat-field frames.  
The individual frames, that were obtained following a raster scheme to improve the final image quality, were then aligned and finally combined. 

\subsection{Archival imaging}

A Gemini Multi-Object Spectrograph (GMOS), 8.1m Gemini North (GN) H$\alpha$ image of CK\,Vul was obtained from the Gemini Observatory Archive under program ID. GN-2010A-Q-62 (PI: A. A. Zijlstra). 
The filter, exposure time and spatial resolution are listed in Table~\ref{tab:obs}.
The individual GMOS exposure was used in combination with the NOT optical exposures to create a composite hybrid image in order to increase the faint nebular emission of Figure \ref{fig:nov}-b.

\subsection{MES high-dispersion spectroscopy}

Long-slit{\color{red},} high-dispersion spectroscopic observations of DQ\,Her were obtained using the Manchester Echelle Spectrometer \citep[MES]{2003RMxAA..39..185M} mounted on the 2.1m telescope at the Observatorio Astron\'{o}mico Nacional in San Pedro M\'{a}rtir (OAN-SPM, Mexico)\footnote{The Observatorio Astronómico Nacional at the Sierra de San Pedro Mártir (OAN-SPM) is operated by the Instituto de Astronomía of the Universidad Nacional Autónoma de México.}. 
The observations were conducted in 2018 in two runs. The first set of spectra were acquired on May 6th and the second on July 23th (see Table~\ref{tab:slits}). 
We used the E2V 42-40 CCD with a pixel size of 13.5 $\mu$m pix$^{-1}$ and a 2$\times$2 on chip binning which resulted in a plate scale of 0.351 arcsec pix$^{-1}$. Observations were made in the spectral range including the H$\alpha$ emission line using a filter ($\Delta\lambda$=90\,{\AA}) to isolate the 87th order (0.05\,{\AA} pix$^{-1}$ spectral scale). The slit width of 150 $\mu$m (1.9 arcsec) corresponds to a spectral resolution of $\simeq$12 $\pm$ 1 km~s$^{-1}$. A mean spatial resolution of 1.3 arcsec was achieved.

The spectra were also processed using a standard calibration routines in {\sc iraf}, including bias subtraction and wavelength calibration with ThAr arc lamps obtained  immediately before and after the science observation.  
The wavelength accuracy is estimated to be $\pm$1 km~s$^{-1}$.

\begin{figure*}
\begin{center}
    \includegraphics[angle=0,width=0.7\linewidth]{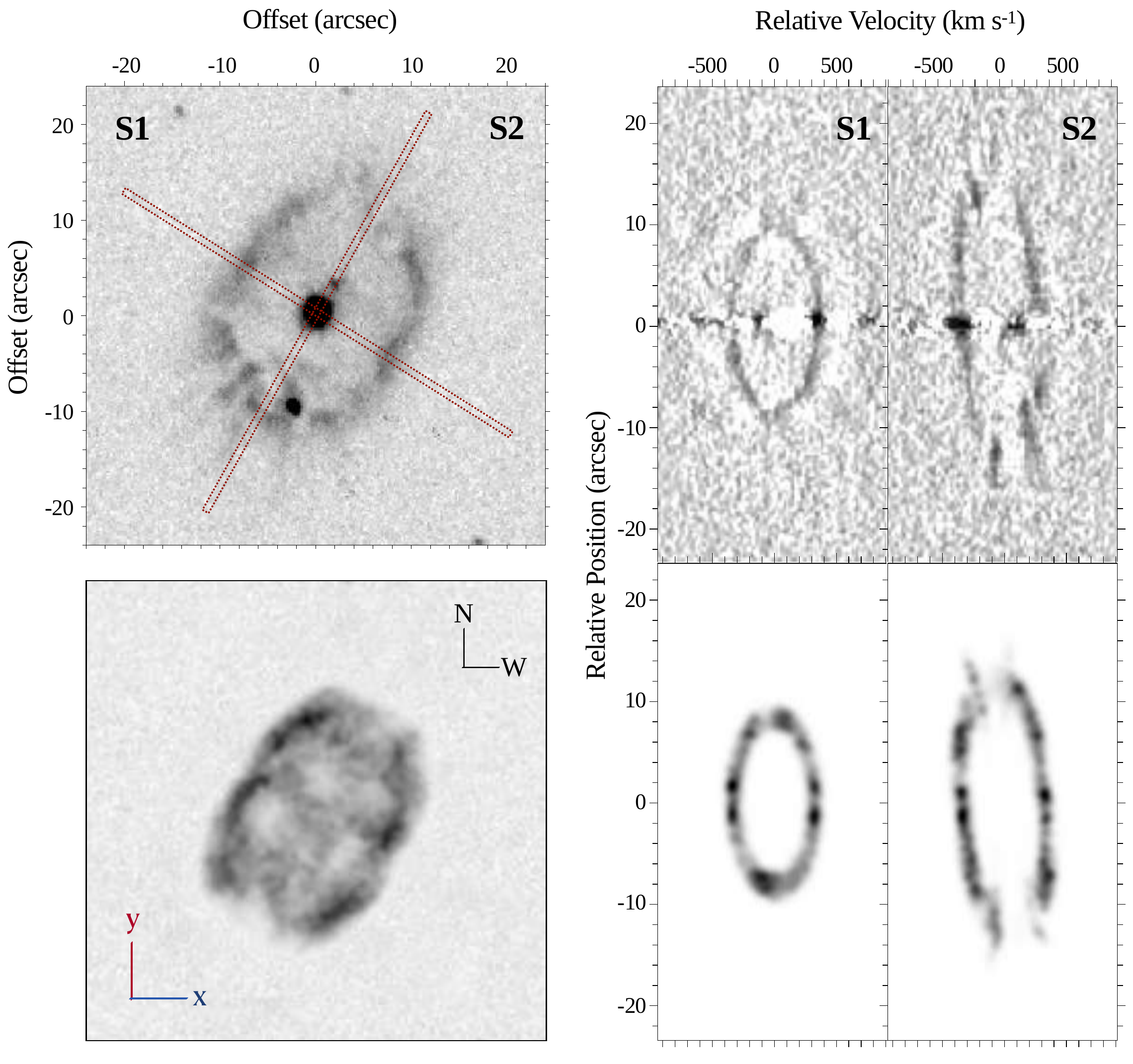}
\caption{
T\,Aur observed (top) and {\sc shape} synthetic (bottom) narrow-band H$\alpha$ image (left) and H$\alpha$ PV maps (right). 
The PV maps were obtained with long-slits along the minor and major axes of T\,Aur, labeled on the direct image as S1 and S2, respectively. 
The stellar continuum has been subtracted from these PV maps. 
The details of the {\sc shape} models are described in the text in Section~\ref{TAur_sec} and Appendix~A. 
}
\label{fig:2}
\end{center}
\end{figure*}

\subsection{ALFOSC intermediate-dispersion spectroscopy}

Intermediate-dispersion{\color{red},} long-slit spectra were conducted with ALFOSC at the NOT of the ORM during three observing runs: on 2018 June, 2020 January, and 2020 July (see Table~\ref{tab:slits}). The same detector used for imaging was used here. We used the ALFOSC grism \#17 with 2400 VPH rules mm$^{-1}$, giving a dispersion of 0.26 \AA~pix$^{-1}$ and covering the 6330-6870 \AA\ spectral range. 
The resulting spectral resolution is 0.65 \AA\ and 1.3 \AA\ for the 0.5 arcsec and 0.9 arcsec slit widths, respectively. The seeing during the observations was $\simeq$0.8 arcsec. 

The spectra were reduced following standard procedures for long-slit spectroscopy within the {\sc iraf} packages. The reduction included bias subtraction, flat-field correction, wavelength calibration and sky subtraction. The wavelength calibration accuracy was $\approx$80 km s$^{-1}$.

In this work, we also used low-resolution spectra obtained with ALFOSC grism \#7 for position S2 in CK\,Vul (Table~\ref{tab:slits}). These observations cover the 3650--7110 \AA\ wavelength range at a spectral resolution of $\simeq$5.8 \AA\,, i.e., $\approx$260 km s$^{-1}$ at the [N~{\sc ii}] $\lambda$6583 \AA\ emission line. The spectra was processed using standard {\sc iraf} routines.

\section{Results}\label{3}

The typical expansion velocity of nova shells \citep[$\gtrsim$500 km s$^{-1}$;][]{2010AN....331..160B} allows us to investigate their kinematics at the spectral dispersion in the range between 12 and 60 km~s$^{-1}$ provided by the MES and ALFOSC observations presented here. Each slit position indeed produces a position-velocity (PV) diagram that reveals the expansion of the nova remnants along the long-slit. This information will be used in conjunction with the direct images to investigate the 3D physical structure of the nova remnant using the morpho-kinematic modelling tool {\sc shape}\footnote{Available from \url{http://www.astrosen.unam.mx/shape/}} \citep[version 5.0;][]{2017A&C....20...87S}. This software allows constructing 3D model structures and producing synthetic images and PV maps under the assumption of homologous expansion that can be compared with the observed ones. 

The description of the observed images, PV maps, and the {\sc shape} models that reproduce them are presented in the following for each nova remnant. 
The details of the {\sc shape} 3D models are presented in Appendix~A. 

\begin{figure*}
\begin{center}
    \includegraphics[angle=0,width=0.7\linewidth]{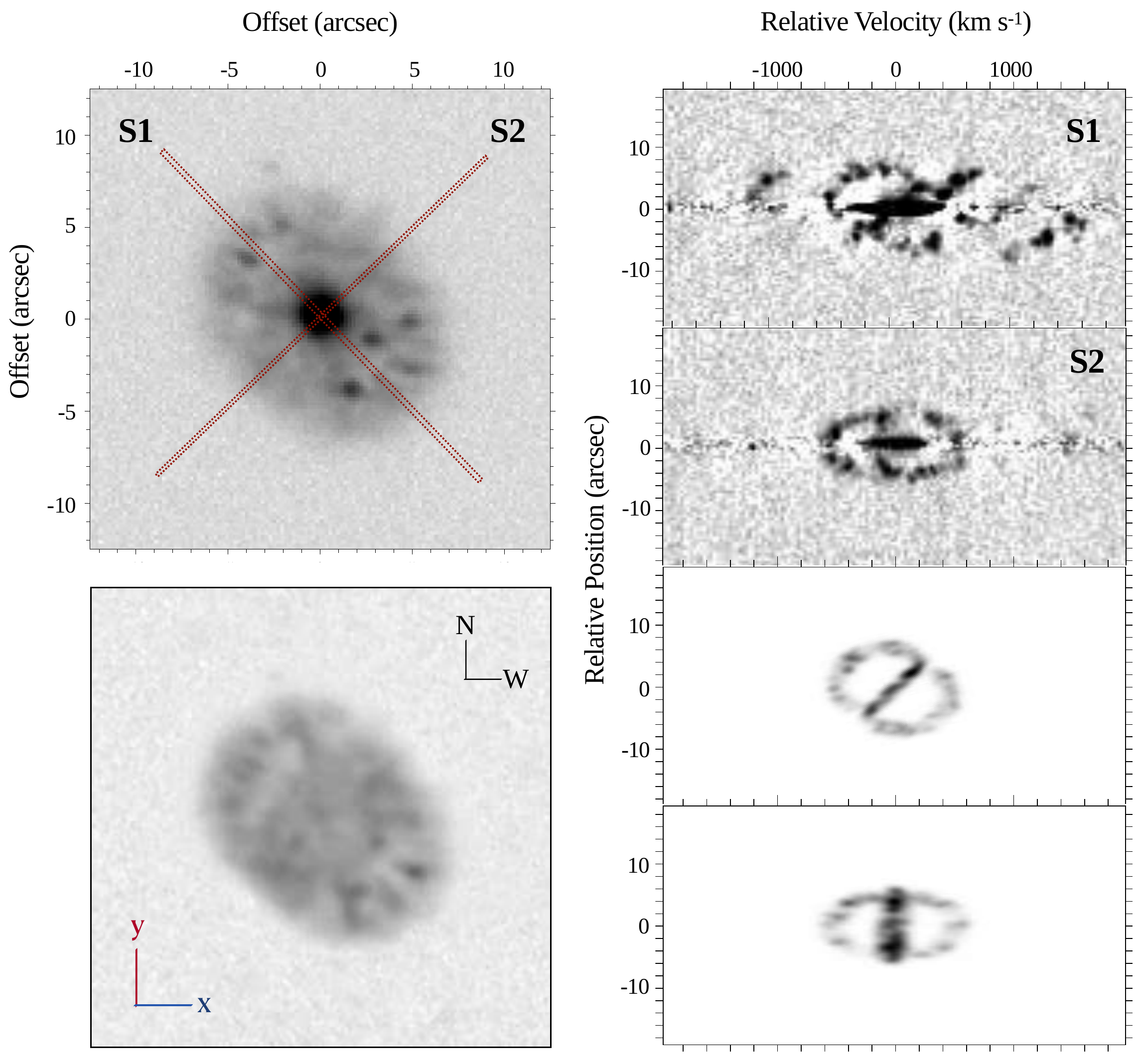}
\caption{
Observed (top) and {\sc shape} synthetic (bottom) narrow-band H$\alpha$ image (left) and H$\alpha$ PV maps (right) of HR\,Del. 
The PV maps were obtained with long-slits along the major and minor axes of HR\,Del, labeled on the direct image as S1 and S2, respectively. 
Note that the PV map along the major axis (S1) includes emission both from the H$\alpha$ and [N~{\sc ii}] $\lambda\lambda$6548,6584 emission lines. 
The stellar continuum is subtracted from these PV maps, but a broad stellar H$\alpha$ emission line is quite noticeable. 
The details of the {\sc shape} models are described in the text in Section~\ref{HRDel_sec} and Appendix~A. 
}
\label{fig:3}
\end{center}
\end{figure*}

\subsection{T\,Aur}
\label{TAur_sec}

T\,Aur experienced a nova event in December 1891. Historically, it was the first nova to be observed spectroscopically \citep{1950PA.....58...50M}, being one of the most observed novae over time \citep{1980ApJ...237...55G, 1995MNRAS.276..353S, 2020ApJ...892...60S}. Its brightness peak reached a $V$ magnitude of 4.5 mag \citep{2010AJ....140...34S}, with a decline time $t_3$ of 100 days \cite[i.e., the time it takes to decline 3 mag from the peak brightness{; see}][]{1957gano.book.....G}. This defines it as a moderately fast nova.

Its H$\alpha$ image in Figure~\ref{fig:2} (top-left panel) shows a shell-like morphology with bright emission along an elliptical ring with multiple clumps.  
These clumps exhibit tails that protrude from the nebular shell, similar to structures that will be described later on in DQ\,Her. The major and minor axes of T\,Aur have angular sizes of 25.4 and 18.6 arcsec, respectively.  

High resolution spectroscopic observations were obtained along the minor and major axes and the corresponding PV maps are shown in the top-right panel of Figure~\ref{fig:2}. The slit along the minor axis shows a structure with lenticular shape, whereas the slit along the major axis indicates a structure open at the tips and clearly tilted. These PV maps thus imply that the nova shell can be basically described as a prolate ellipsoid tilted along the major axis.  

The final nova shell model of T\,Aur deviated slightly from such a prolate ellipsoid, as it rather has a peanut shape (bottom panel of Fig.~\ref{fig:2} and Fig.~\ref{fig:fig8}) with its major axis inclined 75$\degr\pm$2$\degr$ with respect to the line of sight. The comparison between the synthetic images and PV maps indeed shows a notable agreement with the observed data (Fig.~\ref{fig:2}). The expansion velocities in the polar and equatorial directions are 480 km~s$^{-1}$ and 350 km~s$^{-1}$, respectively, whilst the systemic velocity is $V_{\rm LSR}$=15.7$\pm$1.2 km~s$^{-1}$. The axial ratio\footnote{The axial ratio is defined as the ratio of the semi-major axis over the semi-minor axis.} of the model is $\simeq$1.5, which is very similar to the observed one given the large inclination of the major axis with the line of sight. 

\subsection{HR\,Del}
\label{HRDel_sec}

HR\,Del is a peculiar slow nova with a declining rate $t_3$ of 250~days. 
Discovered in 1967, it reached a $V$ maximum peak of brightness at outburst of 3.8 mag. The first image resolving the shell obtained in the emission line of [O\,{\sc iii}] showed an oval remnant with size 3.7$\times$2.5 arcsec \citep{1981MNRAS.196P..87K}. Later on, the deconvolution of ground-based narrow-band images implied a ring-like shape in the H$\alpha+$[N\,{\sc ii}] filter, but a more elongated, bipolar structure in [O\,{\sc iii}] \citep{1994MNRAS.266L..55S}. Finally, \citet{2003MNRAS.344.1219H} obtained \emph{HST} WFPC2 narrow-band images that revealed that the ejecta of HR\,Del consists of a large number of compact knots distributed in a bipolar structure and an equatorial ring.  

\begin{figure*}
\begin{center}
    \includegraphics[angle=0,width=0.7\linewidth]{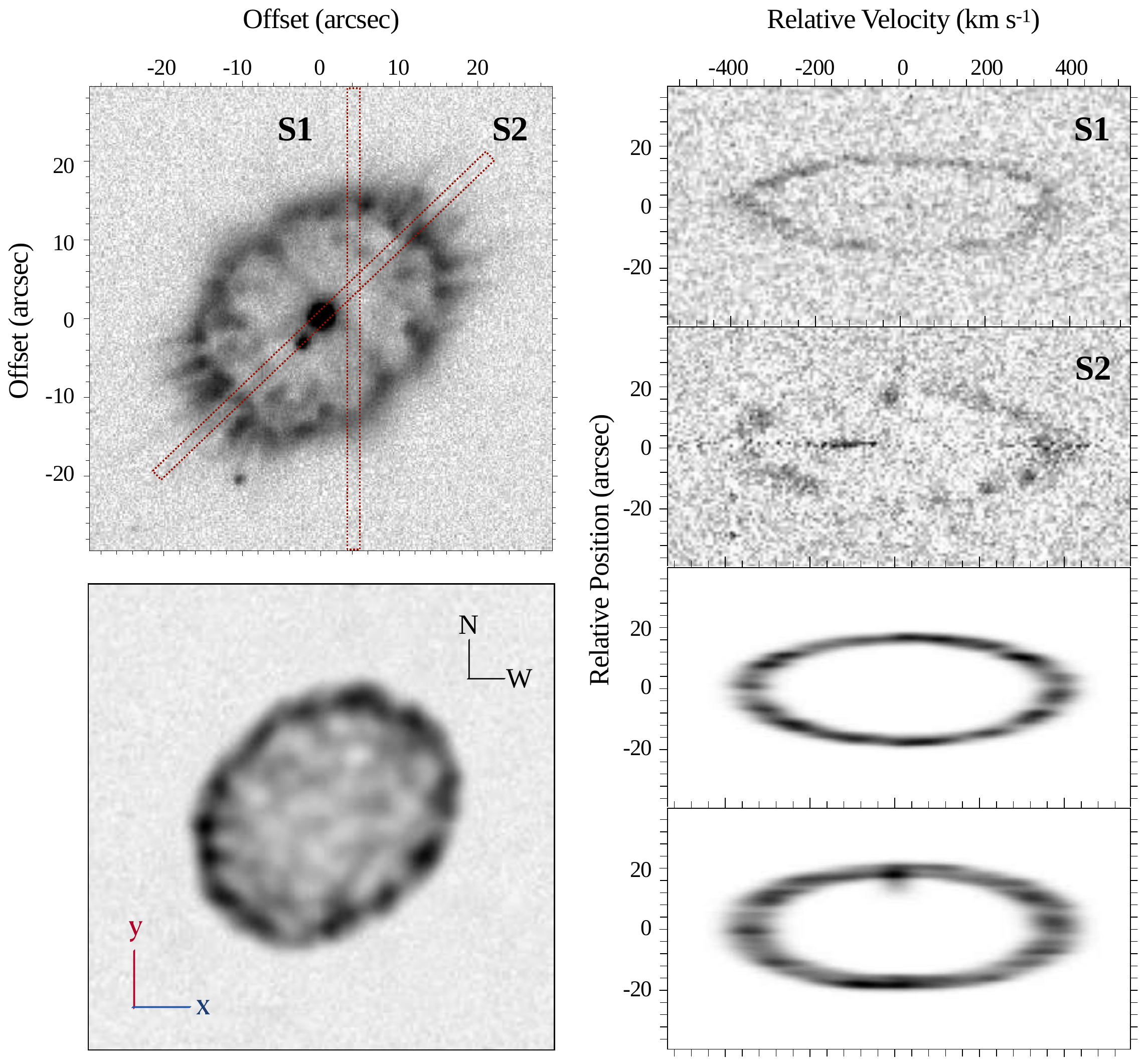}
\caption{
Observed (top) and {\sc shape} synthetic (bottom) narrow-band H$\alpha$ image (left) and H$\alpha$ PV maps (right) of DQ\,Her. 
The PV maps were obtained with long-slits along a PA=0$^\circ$ close to the central star and along the major axis, labeled on the direct image as S1 and S2, respectively. 
The stellar continuum is subtracted from these PV maps. 
The details of the {\sc shape} models are described in the text in Section~\ref{DQHer_sec} and Appendix~A. 
}
\label{fig:dq}
\end{center}
\end{figure*}

The NOT ALFOSC H$\alpha$ image of HR\,Del presented in the top-left panel of Figure~\ref{fig:3} shows an elliptical shell with a clumpy morphology. 
The knots, as noted by \citet{2003MNRAS.344.1219H}, are especially concentrated along the major axis oriented at PA $\simeq$45$^\circ$.  
The size of the source is 13.7$\times$10.3 arcsec. 

Spectroscopic observations were obtained using long-slits aligned along the major and minor axes. The corresponding PV maps are presented in the right panels of Figure~\ref{fig:3}. These include emission from the H$\alpha$ and [N~{\sc ii}] $\lambda\lambda$6548,6584 emission lines for the slit S1 along the major axis, but only H$\alpha$ emission is detected in the slit S2 along the minor axis. 
As in the case of T\,Aur, the slit along the minor axis has a lenticular shape without tilt, whereas that along the major axis shows an obvious line tilt. The PV maps also show a bright central band and a subtle waist for the line profile.  

The morphology of the PV maps suggests that the remnant shell can be interpreted as a prolate ellipsoid tilted along its major axis with respect to the line of sight, but the bright central band is also indicative of an additional equatorial structure. We present a model along these lines in Appendix~A.   
The inclination of the major axis of the shell with to the line of sight and the systemic velocity are estimated to be 37$\degr\pm$3$\degr$ and $V_{\rm LSR}$=43.1$\pm$4.3 km~s$^{-1}$, respectively. The polar expansion velocity is 615 km~s$^{-1}$, whereas the equatorial expansion velocity is 360 km~s$^{-1}$.  

The expansion velocities for HR\,Del reported by several authors range between 520-560 km~s$^{-1}$ \citep{1981MNRAS.196P..87K,1983ApJ...268..689C,1983ApJ...273..647S,2003MNRAS.344.1219H,2009AJ....138.1541M}, being consistent with our determination of the expansion velocity. The aspect ratio of the model in Appendix~A is $\simeq$1.5. 

\subsection{DQ\,Her}
\label{DQHer_sec}

DQ\,Her (Figure \ref{fig:nov}-d) was detected on 1934, reaching a maximum apparent magnitude of 1.4 mags. It is a moderately fast nova that declined with a $t_3$ of 94 days. DQ\,Her is one of the most studied classical novae \citep[and references therein]{1981ApJ...244.1022F, 1986ApJ...308..765S, 1993ApJ...410..357H} and has been suggested to be the prototype of the dust-forming class of novae and the archetype of intermediate polar binary systems \citep{1994PASP..106..209P}. 
Recently, \citet{2020MNRAS.495.4372T} described the presence of extended X-ray emission filling the nova shell and associated with a bipolar magnetized jet, which would be the first one detected in a nova. 

The H$\alpha$ image of DQ\,Her in Figure \ref{fig:dq} exhibits an elliptical symmetry with a slightly pinched central ring. The brightest emission is concentrated in this elliptical shell, with major and minor dimensions and angular expansions of 32.0$\times$24.2 arcsec and 0.19$\times$0.14 arcsec~yr$^{-1}$, respectively \citep{2020ApJ...892...60S}. The knotty structures and cometary knots produce pronounced tails along the major axis. \cite{1995MNRAS.276..353S} argued that these components originate from a fast wind as it breaks through the clumpy shell, whereas \cite{VOR2007} rather supported the presence of a fast ablating stellar wind collimated along the polar axis. 

\begin{figure*}
\begin{center}
    \includegraphics[angle=0,width=0.7\linewidth]{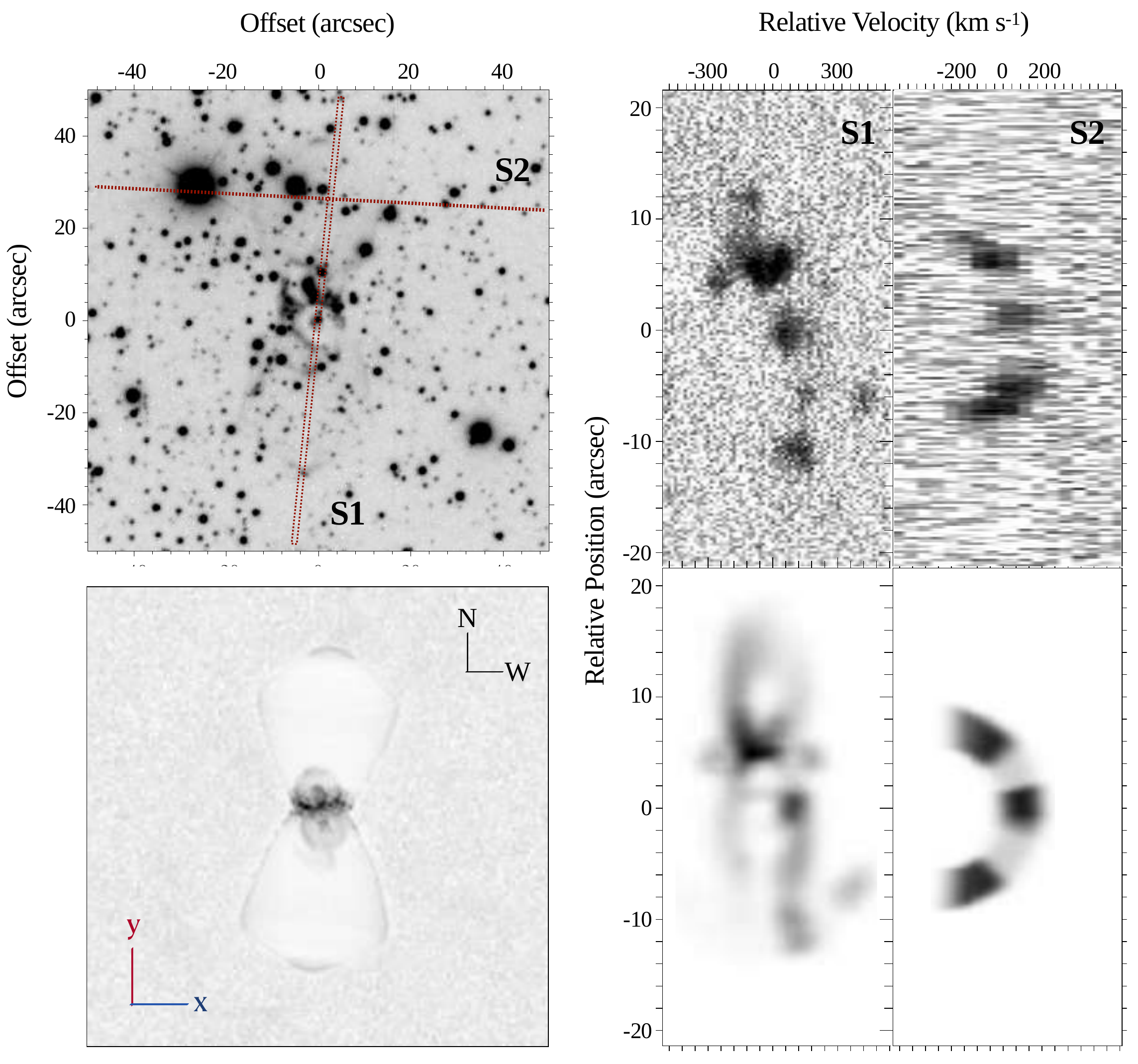}
\caption{
Observed (top) and {\sc shape} synthetic (bottom) narrow-band H$\alpha$ image (left) and H$\alpha$ PV maps (right) of CK\,Vul. 
The PV maps were obtained with long-slits along a PA=174$^\circ$ at the location of the central star (S1) and across the northern bipolar lobe along a PA=88$^\circ$ (S2). 
The details of the {\sc shape} models are described in the text in Section~\ref{CKVul_sec} and Appendix~A. 
}
\label{fig:5}
\end{center}
\end{figure*}

High dispersion spectroscopic observations along a slit at PA=0$^\circ$ close to the central star and along the major axis have been obtained to investigate its kinematics. As revealed by the H$\alpha$ PV maps shown in the right panel of Figure~\ref{fig:dq}, the line profiles have elliptical shapes, with 
a subtle line tilt for the PV map along the major axis. These PV maps therefore imply that the nova shell can be described as a prolate ellipsoid with very little inclination along the major axis. The spatio-kinematic is thus similar to that seen in T\,Aur, but without its break at the tips along the polar direction.

The images and PV maps have been suitably reproduced using {\sc shape} and a prolate ellipsoidal shell. No special attempt has been made to reproduce the remarkable knotty morphology of the shell, as this feature does not affect the basic replication of the observations. We derived a systemic velocity of $V_{\rm LSR}$=$-$16.8$\pm$1.3 km~s$^{-1}$ using the position through the central star. 
The polar expansion velocity of the nova is 460 km~s$^{-1}$, whereas the equatorial expansion velocity is 350 km~s$^{-1}$. The inclination angle is estimated to be 87$\degr\pm$2$\degr$ with respect to the line of sight. Accordingly, the large inclination of the major axis with the line of sight implies that the true axial ratio of $\simeq$1.3 is close to the observed one.  

\subsection{CK\,Vul}
\label{CKVul_sec}

This peculiar object, discovered on 1670, had a maximum brightness of 3 mags at outburst, but its central star has not been detected since its decline in 1972 \citep{1992iue..prop.4466N}. 
Historically, it has been catalogued as an old-slow nova \citep{1982ApJ...258L..41S,1985ApJ...294..271S}, but recently \cite{2020ApJ...904L..23B} proposed that the nature of CK\,Vul outburst belongs to the class of intermediate-luminosity optical transients (ILOTs) known as luminous red novae \citep[LRNe,][]{2002AIPC..637...52M,2019A&A...630A..75P}. 

The H$\alpha$ image of CK\,Vul in Figure~\ref{fig:nov}-b and top-left panel of Figure~\ref{fig:5} displays several components, including a central nebula consisting of a distorted ring-like structure and an inner pair of bipolar lobes, and a more extended pair of bipolar lobes along the north-south direction with bow-shock-like features at their tips. 
The central ring observed in Figure~\ref{fig:5} shows an elliptical morphology with dimensions of 12.2$\times$10.9 arcsec along the polar and equatorial axes, respectively, whereas the inner and outer pairs of bipolar lobes have sizes $\simeq$20$^{\prime\prime}$ and $\simeq$70$^{\prime\prime}$, respectively. 

The spectroscopic data on the top-right panel of Figure~\ref{fig:5} reveals a number of features in the central region with expansion velocities in the range from $-300$ to $+$490 km~s$^{-1}$, which is consistent with the systemic velocity of $-50$ km~s$^{-1}$ and outflow velocities up to 360 km~s$^{-1}$ reported by \citet{2007MNRAS.378.1298H}. Our {\sc shape} model requires the presence of a bipolar structure with an hourglass shape with an extension of 38 arcsec. 
The spectra suggest that the symmetry axis of this hourglass structure is aligned with the plane of the sky. The model additionaly requires an inner bipolar structure with an extension of 22 arcsec (see Fig.~\ref{fig:5} bottom-left panel) nested inside the outer hourglass nebula. This inner bipolar structure, however, is not aligned with the outer bipolar lobes, requiring an inclination angle of 24$\degr\pm$2$\degr$ with respect to the line of sight, which is similar to the 25$\degr$ reported previously \citep[see][and references therein]{2013MNRAS.432..167H}. 
We note that \citet{Kaminski2021} recently reported the presence of these two bipolar structures using optical and molecular emission from ALMA.  
The unprecedented view of the nebula presented there reveal that the molecular emission seems to be encompassed by the inner bipolar structure, which is suggested to be the result of a recent mass ejection with velocities $\sim$470 km s$^{-1}$. 
The evident complex morphological signatures of this nebula suggest a different evolutionary path of its progenitor system than the other objects studied here \citep[e.g.,][]{2018MNRAS.481.4931E,2020A&A...644A..59K,2021arXiv210804305M}.

\subsection{QU\,Vul}
\label{QUVul_sec}

QU\,Vul was discovered on 1984, reaching a maximum brightness at $V$ of 5.6 mag with a decline time $t_3$ of only 49 days \citep{2000AJ....120.2007D}. It is thus one of the fastest novae. The first reports of an optical shell were presented by \cite{VAL1997}, who reported a radius of the shell of 0.95 arcsec in 1994. 

\begin{figure*}
\begin{center}
\includegraphics[angle=0,width=1.0\linewidth]{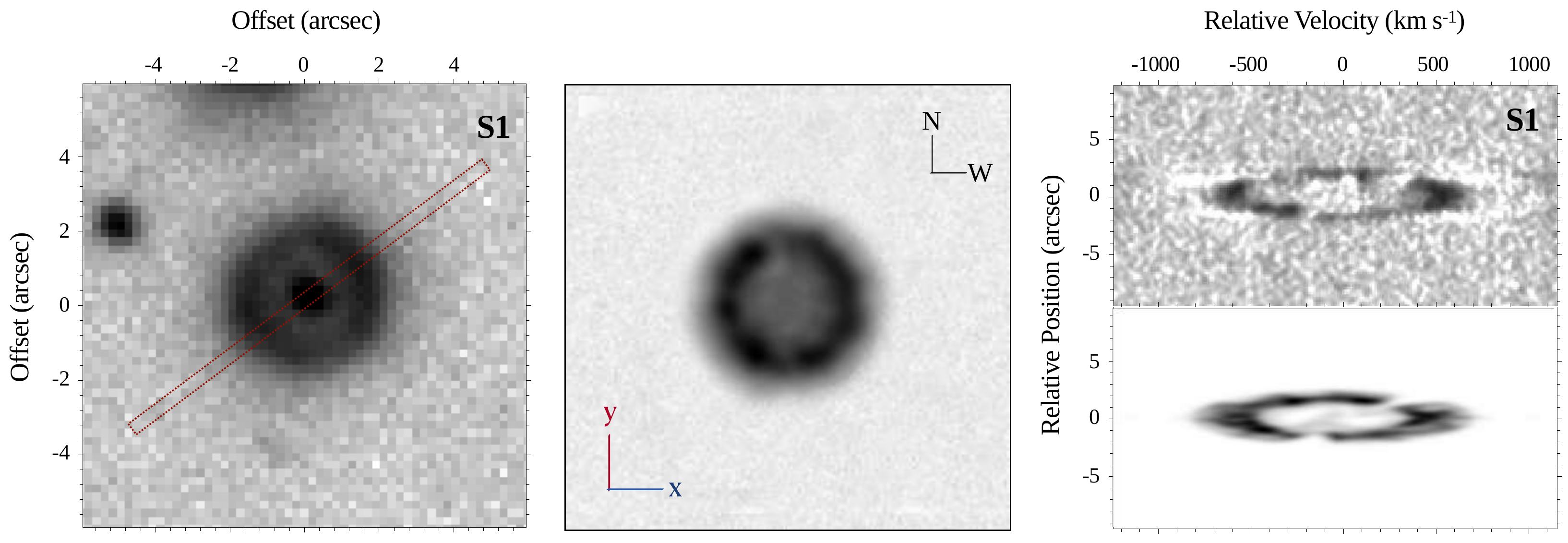}
\caption{
QU\,Vul observed and {\sc shape} synthetic narrow-band H$\alpha$ images (left and center, respectively) and H$\alpha$ PV maps (right-top and right-bottom, respectively). 
The stellar continuum is subtracted from the observed PV map.  
The details of the {\sc shape} models are described in the text in Section~\ref{QUVul_sec} and Appendix~A.
}
\label{fig:6}
\end{center}
\end{figure*}

The morphology of QU\,Vul displayed by its H$\alpha$ image in Figure~\ref{fig:6} is consistent with a spherical shape, with a subtle elongation along the Northwest-Southeast direction that will be hereafter referred as its major axis.
The radius measured from the image is $\simeq$2.1 arcsec, which implies an expansion rate of 0.058~arcsec~yr$^{-1}$. This result is smaller than the one of 0.095 arcsec~yr$^{-1}$ previously reported by \cite{VAL1997} based on an image from an earlier epoch that did not fully resolve the nova shell. It must be noted the notable inhomogeneous brightness  distribution, with the brightest and most well defined regions being located along the direction of the major axis. 

Spectroscopic observations were obtained using a long-slit aligned along its major axis, as labeled on the right panel of Figure~\ref{fig:6}. 
The PV map along this slit shows an elliptical shape with gaps along a few positions and velocity loci of the line profile. 
Since there is no line tilt and the image is consistent with a spherical shell, the nova shell has been modeled in {\sc shape} using a spherical shell.  

The {\sc shape} synthetic image and PV map (middle and bottom-right panels of Fig.~\ref{fig:6}, respectively) derived from a spherical shell provide a reasonable representation of the observed image and PV map.  
The {\sc shape} model implies an expansion velocity in the H$\alpha$ emission line of 660 km~s$^{-1}$ for a systemic velocity $V_{\rm LSR}$=$-$34$\pm$16 km s$^{-1}$. 
This expansion velocity is smaller than those reported in the literature, with values ranging between 790 to 1500 km~s$^{-1}$ \citep{1987Ap&SS.130..157R,1991ANDREA,VAL1997}, but the velocity extent of the PV map definitely validates this lower expansion velocity. 

\begin{figure*}
\begin{center}
\includegraphics[angle=0,width=1.0\linewidth]{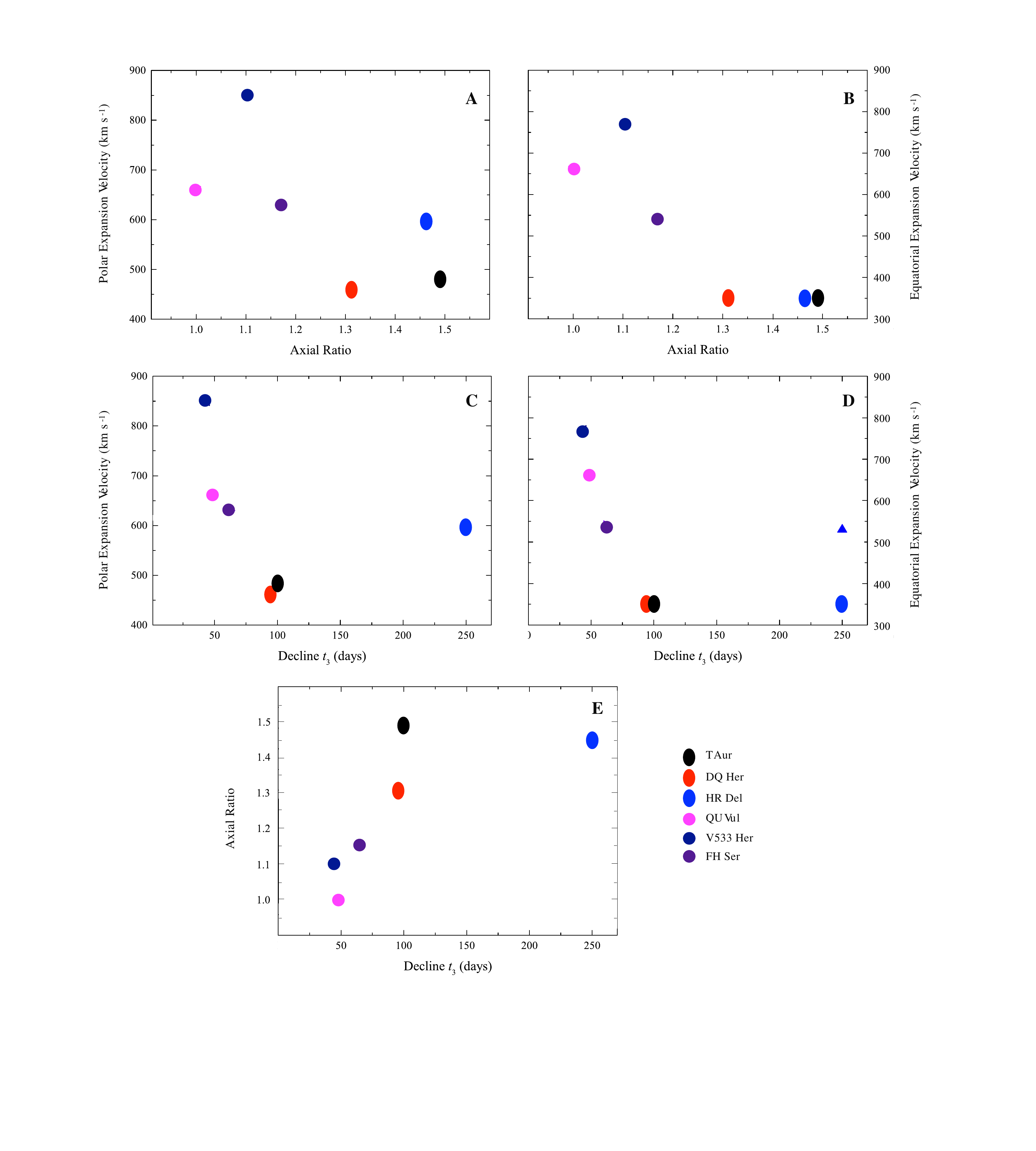}
\caption{
\textbf{Comparison of} the polar (left) and equatorial (right) expansion velocities with the intrinsic axial ratio (top panels) and decline time $t_3$ (middle panels) of nova remnants with suitable spatio-kinematic models \citep[this work][]{}.  
The bottom panel shows the comparison between the intrinsic axial ratio and decline time $t_3$. The circular dots represent nova shells with "spherical" morphologies, whereas the elliptical points correspond to ellipsoidal shells.
}
\label{fig:7}
\end{center}
\end{figure*}

\section{Discussion}\label{4}

The spatio-kinematic models of the nova remnants presented in the previous section disclose their 3D shape. All the nova remnants in our sample can be roughly described by prolate ellipsoids of different aspect ratios, but CK\,Vul with a double bipolar morphology. HR\,Del has an overall ellipsoidal shape requiring an additional clumpy equatorial torus-like structure. We note that there are suitable spatio-kinematic models in the literature for only another two nova shells, namely V533\,Her and FH\,Ser. The former, V533\,Her, can be described as a prolate ellipsoid with aspect ratio 1.1 and polar expansion velocity of 850 km~s$^{-1}$ \citep{VAL1997}, whereas the latter, FH\,Ser, has been described as a prolate ellipsoid with an ellipticity of 1.3 and a polar  expansion velocity 640 km~s$^{-1}$ \citep{2000MNRAS.314..175G}. The present study thus increases significantly the sample of nova remnants with complete spatio-kinematic models of their 3D structure.  

The shaping of the nova shell would occur as the binary companion orbital energy and angular momentum is transferred to the expanding envelope while the ejecta of the nova outburst flows past it. 
As slower novae are expected to have longer interactions with the binary companion, therefore resulting in more highly ellongated remnants than those of faster novae, the asphericity of a nova remnant can be expected to correlate with its expansion velocity and speed class as defined by its decline time $t_3$ \citep{1991ApJ...374..623S,1993PASP..105.1373I}.  
\citet{1995MNRAS.276..353S} conducted an investigation of a sample of 15 nova remnants and indeed concluded that there is a correlation between the axial ratio projected on the sky and $t_3$. However, we note that \citet{1995MNRAS.276..353S} did not take into account projection effects of the 3D structure neither the axial ratio was compared to the true expansion velocity of the nova remnant.  

The correlations between the polar and equatorial expansion velocities, intrinsic axial ratio, and decline time $t_3$ of nova remnants with ellipsoidal morphology analysed here, namely T\,Aur, DQ\,Her, HR\,Del, and QU\,Vul, are shown in Figure~\ref{fig:7}. The correlation coefficients in panels A and B are $-0.66$ and $-0.88$, respectively, whereas those in panels C, D, and E after excluding HR\,Del increase to $-0.92$, $-0.98$, and $+0.93$, respectively. The corresponding information for V533\,Her and FH\,Ser has been also included in these plots. This figure corroborates the anti-correlation between the expansion velocity and intrinsic axial ratio of the nova shell (top panels of Fig.~\ref{fig:7}).  
There is also a strong anti-correlation between the expansion velocity and decline times $t_{3}$ if the nova shell HR\,Del were to be excluded (middle panels of Fig.~\ref{fig:7}). The long decline time of HR\,Del would require an expansion velocity much slower than its measured polar and equatorial velocities.  
On the other hand, there is a clear correlation between the intrinsic axial ratio and the decline times $t_{3}$ (bottom panel of Fig.~\ref{fig:7}), once again if HR\,Del were to be excluded.  

The peculiarities of HR\,Del among the sample of ellipsoidal nova shells is outstanding. In particular, its high expansion velocities along the equator of 360 km~s$^{-1}$ and along the poles of 615 of km~s$^{-1}$ would imply a short decline time $t_3$ inconsistent with the observed value of 250 days. It is also the only ellipsoidal nova shell with [N~{\sc ii}] emission in addition to the H$\alpha$ emission. Furthermore its physical structure as revealed by Balmer and forbidden emission lines is notably different too \citep{2009AJ....138.1541M}. 

\section{Summary}\label{5}

We presented the analysis of optical images and high-resolution spectra of 4 nova systems (T\,Aur, HR\,Del, DQ\,Her and QU\,Vul) and a nova-like object (CK\,Vul). We interpreted our observations making use of the {\sc shape} software to produce morpho-kinamatic models for each object. 

Most of these nova remnants are elliptical but with different degrees of eccentricity: from almost a spherical shell in the case of QU\,Vul to the peanut-shaped shell of HR\,Del. Although we also modelled CK\,Vul, we note that this target has been widely discussed in the past to not belong to classic nova systems. Our {\sc shape} model of this object requires an hourglass outer structure with a nested bipolar structure inside similar as found by other authors. 

The resultant 3D spatio-kinematic {\sc shape} models unveil the structures of the nebulae around these systems, giving us the opportunity of estimating their true axial ratios. The fastest expanding novae with shortest decline times $t_{3}$ also have the smallest axial ratios, that is, their morphologies are close to a spherical shell. Therefore, the kinematic information and the true axial ratio of the nova remnants presented here confirm previous suggestions that more spherical nova shells correspond to fast novae events whilst more elliptical shells are formed by slow nova events, unambiguously confirming the link between the nova asphericity and its formation mechanism.  

\section*{Acknowledgments} 

E.S.\ acknowledges support from Universidad de Guadalajara and
Consejo Nacional de Ciencia y Tecnolog\'{i}a (CONACyT) for a student scholarship. M.A.G.\ acknowledges support of grant PGC 2018-102184-B-I00 of the Ministerio de Educación, Innovación y Universidades cofunded with FEDER funds.
S.Z.\ acknowledges support from (TecNM) 11189.21-P, from F.Ramos-DIEE/MIA/ITE, E. Bugarin-DIEE/DEPI/ITE and J. Moreno-DIEE/MIA/ITE. G.R.-L.\ acknowledges support from Universidad de Guadalajara, CONACyT grant 263373 and Programa para el Desarrollo Profesional Docente (PRODEP, Mexico). J.A.T.\ acknowledges funding from the Marcos Moshinsky Foundation (Mexico) and Dirección General de Asuntos del Personal Académico (DGAPA), Universidad Nacional Autónoma de México, through grants Programa de Apoyo a Proyectos de Investigación e Innovación Tecnológica (PAPIIT) IA100720. L.S.\ acknowledges support from PAPIIT grant IN101819. The data presented here were obtained in part with ALFOSC, and provided by the Instituto de Astrofísica de Andalucía (IAA) uder a joint agreement with the University of Copenhagen and NOTSA. This paper is based partly on ground-based observations from the Observatorio  Astron\'{o}mico Nacional at the Sierra de San Pedro M\'{a}rtir (OAN-SPM), which is a national facility operated  by the Instituto de Astronom\'{i}a of the Universidad Nacional Aut\'{o}noma de M\'{e}xico. Based on observations obtained at the international Gemini Observatory, a program of NSF’s NOIRLab, which is managed by the Association of Universities for Research in Astronomy (AURA) under a cooperative agreement with the National Science Foundation on behalf of the Gemini Observatory partnership: the National Science Foundation (United States), National Research Council (Canada), Agencia Nacional de Investigaci\'{o}n y Desarrollo (Chile), Ministerio de Ciencia, Tecnolog\'{i}a e Innovaci\'{o}n (Argentina), Minist\'{e}rio da Ci\^{e}ncia, Tecnologia, Inova\c{c}\~{o}es e Comunica\c{c}\~{o}es (Brazil), and Korea Astronomy and Space Science Institute (Republic of Korea).

\section*{Data Availability} 

The data underlying this work are available in the article.
The data files will be shared on request to the first author.








\appendix

\section{3D models of novae}

\begin{figure*}
\begin{center}
\includegraphics[width=0.9\linewidth]{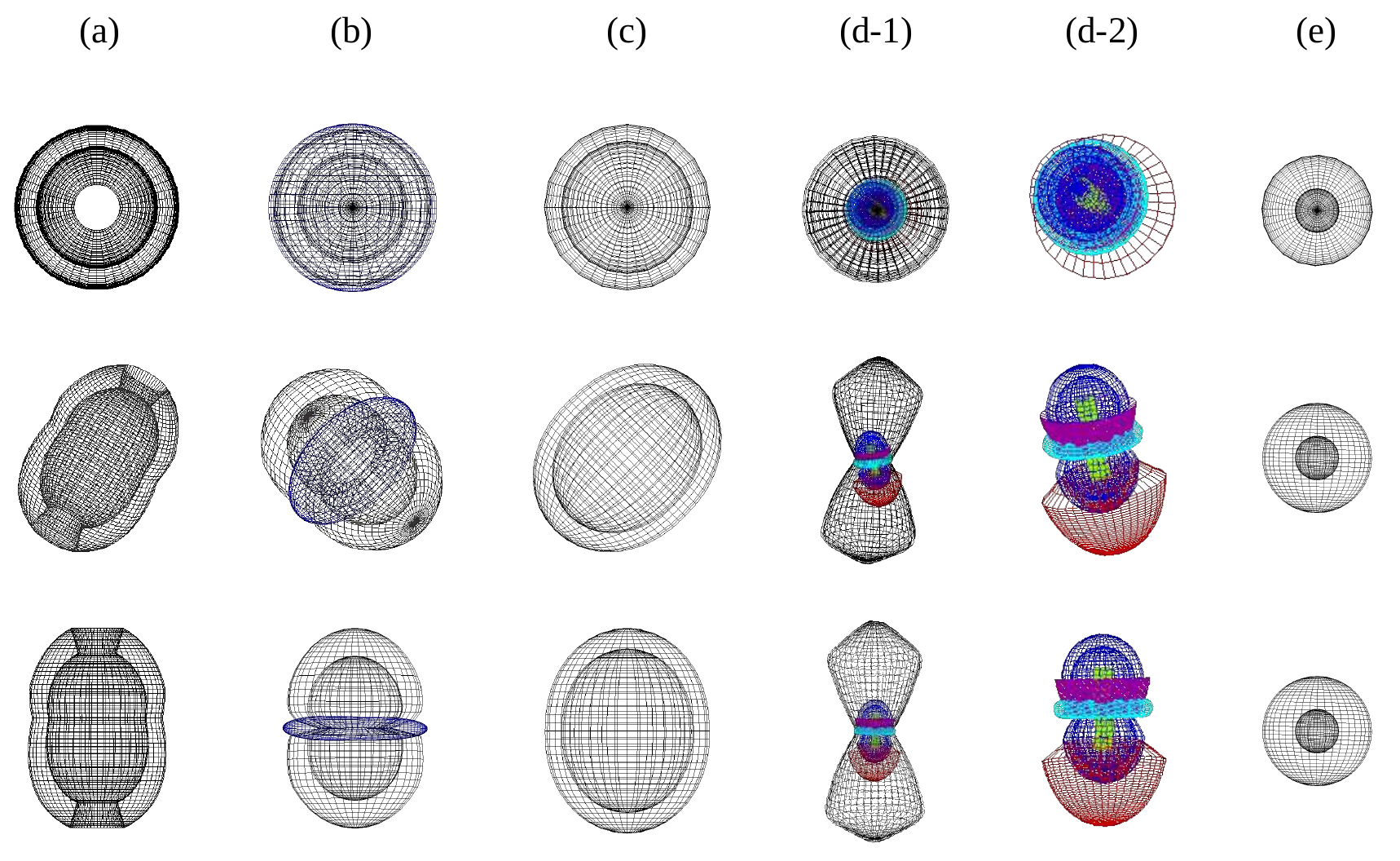}
\caption{{\sc shape} mesh models of the novae, a) T\,Aur, b) HR\,Del, c) DQ\,Her, d-1) CK\,Vul, d-2) center of CK\,Vul and e) QU\,Vul. These models are rendered at three orientations: top-view (first row), projected-view (second row) and side-view (third row). The expansion is assumed to be homologous.
}
\label{fig:fig8}
\end{center}
\end{figure*}

Images and high-resolution spectra in this paper were used in order to construct 3D models of four nova shells and one nova-like source using the spatio-kinematic code {\sc shape}. Figure \ref{fig:fig8} shows the main structure of each object in different orientations. The main geometry of T\,Aur, HR\,Del and DQ\,Her can be modeled by ellipse, detailing the representative characteristics of each of them (peanut, bipolar and prolate shaped shells, respectively). In the case of CK\,Vul the internal structure has been separated for improved visualization and is displayed in column d-2. Finally, QU\,Vul was modeled with a spherical shape and some internal features to represent the gaps seen in the observed spectrum.

\bsp	
\label{lastpage}
\end{document}